# Simple piezoelectric-actuated mirror with 180 kHz servo bandwidth

**Travis C. Briles, Dylan C. Yost, Arman Cingöz, and Jun Ye**

*JILA, National Institute of Standards and Technology and University of Colorado*
*Department of Physics, University of Colorado, Boulder, Colorado 80309-0440, USA*

**Thomas R. Schibli**

*Department of Physics, University of Colorado, Boulder, Colorado 80309-0440, USA*
*travis.briles@colorado.edu*

**Abstract:** We present a high bandwidth piezoelectric-actuated mirror for length stabilization of an optical cavity. The actuator displays a transfer function with a flat amplitude response and greater than 135° phase margin up to 200 kHz, allowing a 180 kHz unity gain frequency to be achieved in a closed servo loop. To the best of our knowledge, this actuator has achieved the largest servo bandwidth for a piezoelectric transducer (PZT). The actuator should be very useful in a wide variety of applications requiring precision control of optical lengths, including laser frequency stabilization, optical interferometers, and optical communications.

## 1. Introduction

Laser systems actively stabilized against phase and frequency fluctuations [1, 2, 3, 4, 5, 6] have become essential tools for a variety of applications such as optical atomic clocks [7], high resolution spectroscopy [8, 9], optical frequency synthesis [10, 11], coherent optical communications [12, 13, 14, 15, 16], and the search for gravitational waves [17]. One of the most common type of actuator used to stabilize an optical system is a mirror mounted on a servo-controlled piezoelectric transducer (PZT), which changes an optical frequency or phase by adjusting a relevant optical path length. Traditionally, PZT actuators have been used for low bandwidth feedback ($\approx$ 5 kHz) in conjunction with a high bandwidth actuator such as an electro-optic modulator (EOM) or an acousto-optic modulator (AOM) [18]. Whereas such configurations are able to achieve very large servo bandwidths, AOMs and EOMs could limit the overall system power, add complexity, and introduce dispersion which must be carefully controlled in femtosecond laser systems. Hence, a high bandwidth PZT-based system is desirable.

When utilized in a servo loop, the bandwidth of PZT actuators is typically limited by strong mechanical resonances between 20 and 40 kHz [19, 20]. Mechanical resonances in the actuator response are problematic since the amplitude response is usually accompanied by significant phase shifts, which can lead to positive feedback in the vicinity of the resonance, causing instabilities in the lock. Therefore, the ideal actuator would have a flat amplitude and phase response over the entire bandwidth of the feedback loop so that the lock performance is completely determined by the servo controller.

Here we present a simple actuator with a flat amplitude and phase response to frequencies above 200 kHz, followed by small resonances that minimally affect lock stability. This actuator outperforms previously reported high performance actuators that rely on complicated designs involving multiple PZTs and highly engineered mechanical systems [21]. The closed loop behavior of the actuator is demonstrated by locking an external Fabry-Perot cavity to a continuous wave (CW) laser with a Pound-Drever-Hall (PDH) scheme [22] where a unity gain frequency of ≈180 kHz is achieved. The actuator should be of wide utility in a variety of applications including the stabilization of both laser frequency and optical interferometers.

## 2. Design Principles

A typical PZT actuator consists of a mirror attached to a PZT, which is itself affixed to a massive mounting structure to absorb shocks. To obtain an actuator with a flat response, each of the constituent parts, including the adhesive, should be chosen to increase the frequency and minimize the quality factor (Q) of any mechanical resonances. In our actuator, we push the resonances to higher frequencies by choosing rigid materials with a high speed of sound and by using a small mirror and PZT. Normal modes of the mounting structure are damped by filling the mounting structure with lead, and the intrinsic piezoelectric resonances are damped by introducing irregularity to the PZT surface. In the following subsections, we present detailed discussions of these design principles. A schematic of the final design for the mounting structure is shown in Fig. 1 (a), and a photograph of the complete actuator in a mirror mount is shown in Fig. 1 (b).

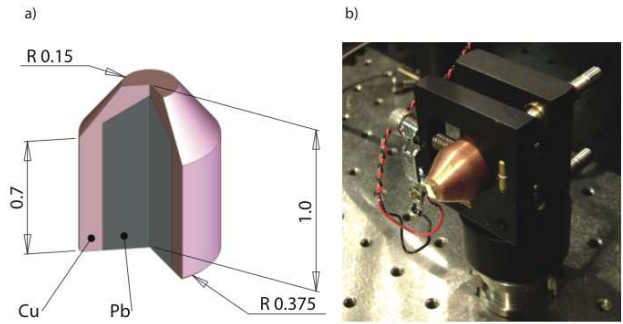

Fig. 1. Schematic of the PZT mounting structure that was used for the experiments described here. Dimensions are in inches. (b) A photograph of the actuator in a mirror mount.

To evaluate different designs, we measured the mechanical response of the actuator at different Fourier frequencies (actuator's transfer function) by placing the actuator in one arm of a Michelson interferometer. For small excursions about the midpoint of the optical fringe, the interference signal is linearly proportional to the change in the relative length between the two arms of the interferometer. The overlapped beam is incident on a photodetector with a flat response over all the frequencies of interest, and the amplitude and phase response of the actuator is then measured with a network analyzer. In addition, we also measure the PZT resonances directly on a network analyzer by measuring the frequency-dependent electrical impedance.

### 2.1. Mounting Structure

Typically, the most troublesome mechanical resonances are longitudinal compression waves of the mounting structure and shearing modes causing drumhead-like vibrations of the mounting face. One approach is to push the resonances to higher frequencies by reducing the physical dimensions of the mounting structure. Unfortunately, this is a poor strategy for longitudinal waves since a typical amplitude for such resonances can be as high as 20 dB, which would require a reduction of the servo gain by an order of magnitude given a typical integral gain slope [23]. In contrast, we have found that tapering the mounting structure so that the front face matches the size of the PZT and the mirror is effective at dealing with drumhead modes, which typically have a much lower Q.

A more effective approach to mitigating the effects of the longitudinal resonances is damping. This is very effectively achieved by drilling out the back of a copper mounting structure and filling it with lead. Such techniques have been used previously in PZT actuators but have not been described [24]. For most effective damping, we have found that the tip of the lead should reach within $\approx$ 1mm of the mounting surface. The benefit of the lead inside the copper mounting structure is clearly indicated in Fig. 2, which compares the transfer functions of two actuators utilizing the same PZT [25], mirror, bonding agent, and the same mounting structures before and after it is filled with lead. The copper mounting structure without lead (bottom curve) shows many sharp resonances that are effectively damped out by the addition of the lead (top curve).

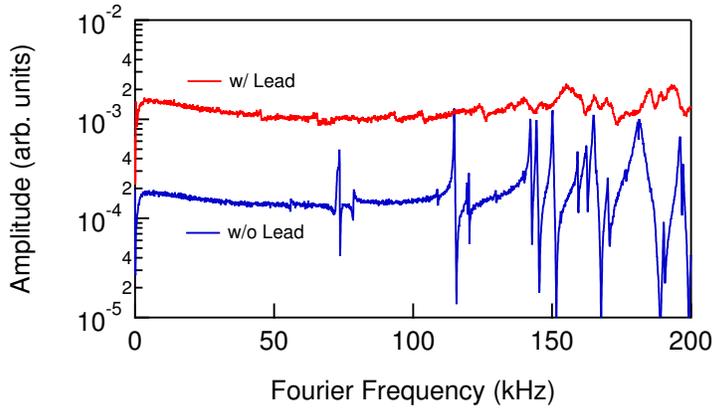

Fig. 2. Amplitude transfer functions for PZT actuators measured with a Michelson interferometer. The actuators were constructed with a stack PZT [25] and a rectangular mirror of the same size mounted on a (bottom curve) copper mount of the same size as described in Fig. 1 and (top curve) the same mount filled with lead. The two curves have been offset for clarity.

*2.2. Mirror and Adhesive*

Insights into the choice of the mirror geometry and adhesive can be gained by modeling the actuator as a system of coupled oscillators where the mirror, PZT, and mounting structure comprise the masses, with the adhesive acting as the spring. The resonant frequencies of the system can be pushed to higher values by increasing the spring constant $k$ or decreasing the mass of the mirror by thinning the mirror substrate. Treating the adhesive as a bar in compression mode gives a spring constant $k = \frac{EA}{L}$, where $A$, $L$ and $E$ are the bar's cross sectional area, length, and Young's Modulus, respectively. In practice, we applied a pressure of $\sim 10^6$ Pa while the rigid (High $E$) adhesive [26] cured to ensure a thin layer. The mirror diameter should also be matched to the size of the PZT to mitigate drumhead vibrations of either the mirror or PZT.

*2.3. Choice of PZT*

Another essential criterion in the design of a high bandwidth PZT actuator is the type and geometry of PZT used [27]. There are two common PZT geometries, tubular and disk. Tubular PZTs are best suited to very low bandwidth ($\sim$ few 100 Hz) applications requiring large travel and the ability to monitor the transmitted light through the PZT-mounted mirror. Disk PZTs are a better choice for high bandwidth actuators because they typically have much higher unloaded resonance frequencies. Disk PZT's are available as either a single piece of piezoelectric ceramic (HV PZT) or as a stack composed of many thin layers of piezoelectric material. HV PZTs have a small sensitivity of 0.5 nm/V, independent of the thickness [28], and hence typically require hundreds of volts to operate. Stack PZTs, on the other hand, have much larger total travel ranges and sensitivities because each layer in the stack moves in concert. Thus, their sensitivity is increased by the number of layers in the stack.

While stack PZT's are very convenient for achieving large displacements ($\sim \mu$m) at low voltages, their large capacitance requires a driver with a high output current in order to reach the voltage slew rates required at high frequencies. With a suitable driver [29], we were able able to achieve a $\sim$ 100 kHz servo bandwidth with a 7.5-mm diameter, $\approx$ 2-mm-thick mirror on a $5 \times 5 \times 2$ mm stack PZT [25] using the mounting structure presented in Fig. 1. While the actuator showed a very flat amplitude response out to 150 kHz, it exhibited pronounced phase roll-off (90° phase margin at 100 kHz).

We have found that HV PZTs show superior phase response over stack PZTs at frequencies above 100 kHz. In addition, simple modifications can be performed on HV PZTs to further improve their phase response. The quality factor for an intrinsic piezoelectric resonance will be the strongest for perfectly flat, parallel surfaces. Thus, the resonances can be damped out by introducing irregularity to the surface of one of the sides. Cutting the PZT's thickness on a diamond saw roughens the surface while simultaneously pushing the resonances to higher frequencies. A major concern during this operation is the depoling of the PZT. In practice, we have found that cutting the PZT slowly leads to no significant change in the sensitivity of PZT, as verified by interferometric measurements.

Intrinsic piezoelectric resonances correspond to maximum conversion of electrical to mechanical energy. The damping effects of surface roughening were evaluated by measuring the frequency dependent impedance as shown in Fig. 3, where the curves have been normalized against the frequency dependence of an equivalent capacitor. The sharp resonances of the commercial PZT with both electrodes intact are effectively damped by scraping off the negative electrode and roughening the surface with an abrasive. Electrical contact was made with copper tape firmly attached to the cut surface.

Actuators were constructed using both an unmodified HV PZT of 0.1 in. thickness and a HV PZT of the same shape that had the negative side cut off on a diamond saw with a final thickness of 0.07 inches. For the PZT without the negative terminal, the cut side was electrically grounded directly to the mounting structure and the phase response for both actuators was recorded with an interferometer as shown in Fig. 4. Although the response functions still show resonances due to a mismatch in the size of the mirror and PZT, it is clear that shortening the PZT and roughening one of its surfaces flatten out the overall phase response. A third actuator made with an uncut but roughened surface PZT showed a phase response in between the two presented in Fig. 4.

## 3. Final Design and Closed-Loop Performance

Our final actuator utilized all of the design principles presented in the preceding section. The actuator consisted of a $3 \times 5 \times 1$ mm mirror and a HV PZT of approximately the same size (cut to $\approx$ 0.1 in. thick) mounted on the mounting structure shown in Fig. 1. The transfer function of this actuator shown in Fig. 5 demonstrates the flat amplitude and phase response with a greater than 135° phase margin past 200 kHz.

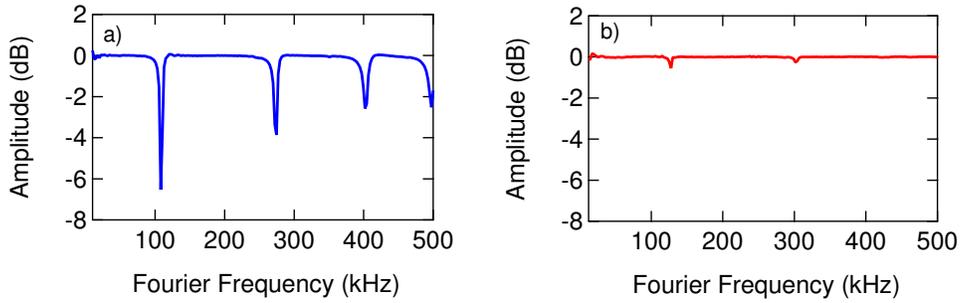

Fig. 3. The frequency-dependent impedance measured with a network analyzer for a commercial HV PZT with both electrodes intact (a) and the same PZT that has had the ground electrode removed and the surface roughed up with an abrasive (b).

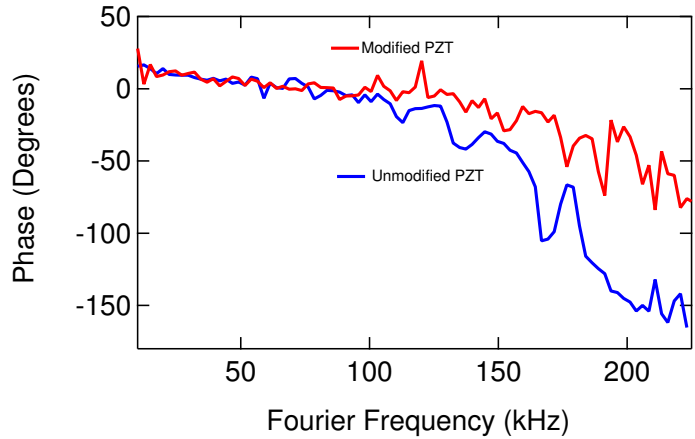

Fig. 4. Interferometric measurement of phase responses for an actuator built with a commercial PZT (0.1 in. thick) and one that has been cut with a diamond saw (0.07 in. thick).

Although the HV actuator presented in Fig. 5 has a much smaller capacitance (on the order of a few nF) than the stack PZT, its response can still be limited due to the finite current output of HV amplifiers. Our HV PZT actuator was driven with both a low current (3 mA), 1-kV amplifier at low frequencies where the need for gain is largest and a high current (50 mA), 200-V amplifier at frequencies above 1 kHz to adequately drive the capacitive load. The outputs of the two amplifiers were combined with a bias tee. The crossover frequency and the gain of the individual amplifiers were adjusted until the voltage at the input to the PZT had a flat response over the bandwidth of the actuator.

The closed loop performance of the actuator was measured by using it to lock the frequency of an external Fabry-Perot cavity to a CW Nd:YAG laser using a PDH feedback [22]. The actuator showed excellent performance, achieving a unity gain frequency of ≈180 kHz and > 40-dB noise suppression at low frequencies, as shown in Fig. 6. Achieving such a high unity gain frequency allows a robust lock with a superior low frequency performance as well as sufficient noise suppression at high frequencies for most applications. While Fig. 6 shows a significant servo bump above 200 kHz, this contributes negligibly to the rms phase noise since the phase noise power spectral density is given by the frequency noise power spectral density weighted by $1/f^2$. This illustrates the importance of a high bandwidth servo loop since the integrated rms phase noise should be roughly inversely proportional to the servo bandwidth.

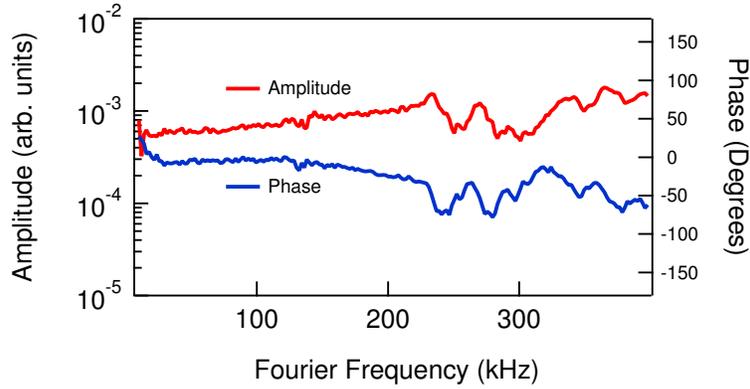

Fig. 5. Transfer function measured with a Michelson interferometer for ∼ 1 mm thick HV PZT disk and ∼1 mm thick mirror on the lead-filled copper mount presented in Fig 1.

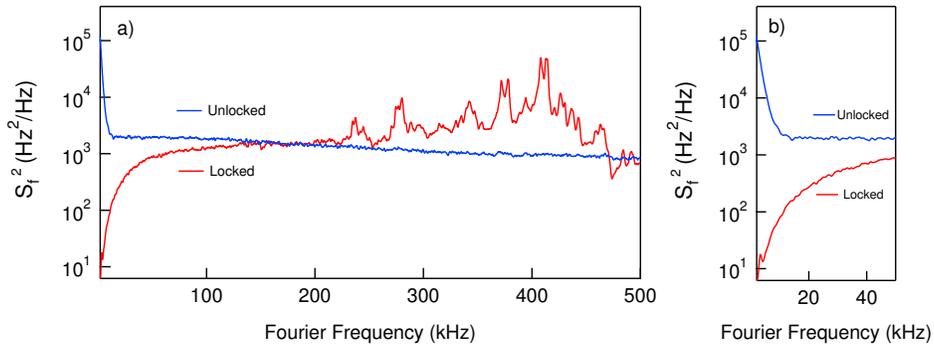

Fig. 6. The converted frequency noise power spectral density of the in-Loop error signal for PDH lock with the PZT described in Fig. 5. In the unlocked case the cavity was manually held on resonance. Panel (a) a full frequency span 0 - 500 kHz, (b) zoomed-in to 50 kHz.

## 4. Conclusion

We have developed a high-bandwidth PZT actuator with superior response functions that should be of wide utility in applications for stabilization of lasers and optical interferometers. The performance relies on the combination of large dynamic range enjoyed by traditional PZTs and high bandwidth typically reserved for actuators such as EOMs and AOMs. The actuator's simple implementation and 180 kHz servo bandwidth removes virtually all of the common laboratory noise such as sound and vibration, making it a superior choice for most applications.

We thank M. Martin, J. L. Hall, A. M. March, and L. Young for helpful discussions. This work was supported by DARPA and NIST. A. Cingöz is a National Research Council postdoctoral fellow. Travis Briles's email is travis.briles@colorado.edu.